\documentclass[preprint,12pt,superscriptaddress]{revtex4-1}

\setcitestyle{super}

\usepackage{graphicx}
\usepackage{amssymb}
\usepackage{epstopdf}
\usepackage{bm}
\usepackage{amsmath}

\makeatletter
\renewcommand\@biblabel[1]{#1.}
\makeatother

\begin{document}
\title{Nanowire spin torque oscillator driven by spin orbit torques}


\author{Zheng Duan}
\thanks{These authors contributed equally to this work.}
\author{Andrew Smith}
\thanks{These authors contributed equally to this work.}
\author{Liu Yang}
\thanks{These authors contributed equally to this work.}
\author{Brian Youngblood}
\thanks{These authors contributed equally to this work.}
\affiliation{Department of Physics and Astronomy, University of California, Irvine, California 92697, USA}
\author{J{\"u}rgen Lindner}
\affiliation{Helmholtz Zentrum Dresden Rossendorf, D-01328 Dresden, Germany}
\author{Vladislav E. Demidov}
\affiliation{Department of Physics and Center for Nonlinear Science, University of M{\"u}nster, Corrensstrasse 2-4, 48149 M{\"u}nster, Germany}
\author{Sergej O. Demokritov}
\affiliation{Department of Physics and Center for Nonlinear Science, University of M{\"u}nster, Corrensstrasse 2-4, 48149 M{\"u}nster, Germany}
\affiliation{Institute of Metal Physics, Ural Division of RAS, Yekaterinburg 620041, Russia}
\author{Ilya N. Krivorotov}
\affiliation{Department of Physics and Astronomy, University of California, Irvine, California 92697, USA}

\begin{abstract}
Spin torque from spin current applied to a nanoscale region of a ferromagnet can act as negative magnetic damping and thereby excite self-oscillations of its magnetization. In contrast, spin torque uniformly applied to the magnetization of an extended ferromagnetic film does not generate self-oscillatory magnetic dynamics but leads to reduction of the saturation magnetization. Here we report studies of the effect of spin torque on a system of intermediate dimensionality -- a ferromagnetic nanowire. We observe coherent self-oscillations of magnetization in a ferromagnetic nanowire serving as the active region of a spin torque oscillator driven by spin orbit torques.  Our work demonstrates that magnetization self-oscillations can be excited in a one-dimensional magnetic system and that dimensions of the active region of spin torque oscillators can be extended beyond the nanometer length scale. 
\end{abstract}

\maketitle

A current of spin angular momentum incident on a ferromagnet exerts torque on its magnetization and drives it out of equilibrium\cite{Slonczewski1996, Berger1996}. Owing to its non-conservative nature, this spin torque (ST) can act as effective negative magnetic damping and thereby excite magnetization self-oscillations\cite{Kiselev2003,Slavin2009}. Spin torque oscillators (STO) have been realized in nanoscale spin valves\cite{Kiselev2003,Ozyilmaz2004,Mistral2006,Braganca2010}, point contacts to magnetic multilayers\cite{Rippard2004,Ruotolo2009,Mohseni2013} and nanoscale magnetic tunnel junctions\cite{Nazarov2006,Deac2008,Houssameddine2008,Georges2009,Rowlands2013}. Recently, a new type of STO based on current-induced spin orbit (SO) torques in a Permalloy(Py)/platinum(Pt) bilayer was demonstrated\cite{Demidov2012,Liu2012,Liu2013}. Spin orbit torques\cite{Haney2013,Kim2012,Martinez2013} in this system can arise from the spin Hall effect in Pt\cite{Dyakonov1971,Hirsch1999,Zhang2000,Ando2008,Fan2013,Hoffmann2013,Bai2013} and the Rashba effect at the Pt/Py interface\cite{Bychkov1984,Sinova2004,Obata2008,Miron2010}.

In all STOs studied to date, the active region where the negative ST damping exceeds the positive Gilbert damping of the ferromagnet was restricted to nanoscale dimensions. A recent study\cite{Demidov2011} of spatially uniform ST applied to an extended ferromagnetic film revealed that coherent self-oscillations of magnetization cannot be excited in this two-dimensional (2D) magnetic system. Instead, spin torque was shown to significantly reduce the saturation magnetization of the film\cite{Demidov2011}. The absence of ST-driven self-oscillations in a 2D ferromagnet was attributed to amplitude-dependent damping arising from nonlinear magnon scattering that prevents any of the multiple interacting spin wave modes of the system from reaching the state of large amplitude self-oscillations. As a result, the energy and angular momentum pumped by ST into the film is redistributed among a large number of spin wave modes leading to reduction of the saturation magnetization of the film. This study raises an important question about the role of the magnetic system dimensionality in ST-induced magnetization dynamics.

In this article, we demonstrate that spatially uniform ST can excite self-oscillations of magnetization in a one-dimensional (1D) magnetic system -- a ferromagnetic nanowire. We report studies of ST-driven dynamics in a Pt/Py bilayer nanowire, in which SO torques excite self-oscillations of magnetization over a 1.8 $\mu$m long active region. This nanowire STO exhibits two types of self-oscillatory modes that arise directly from the edge and bulk spin wave eigenmodes of the Py nanowire \cite{Zheng2014}. Our work suggests that geometric confinement of the spin wave spectrum in the 1D nanowire geometry limits the phase space for nonlinear magnon scattering compared to the 2D film geometry and thereby enables STOs with a spatially extended active region.

\begin{figure}
\includegraphics[width=\linewidth]{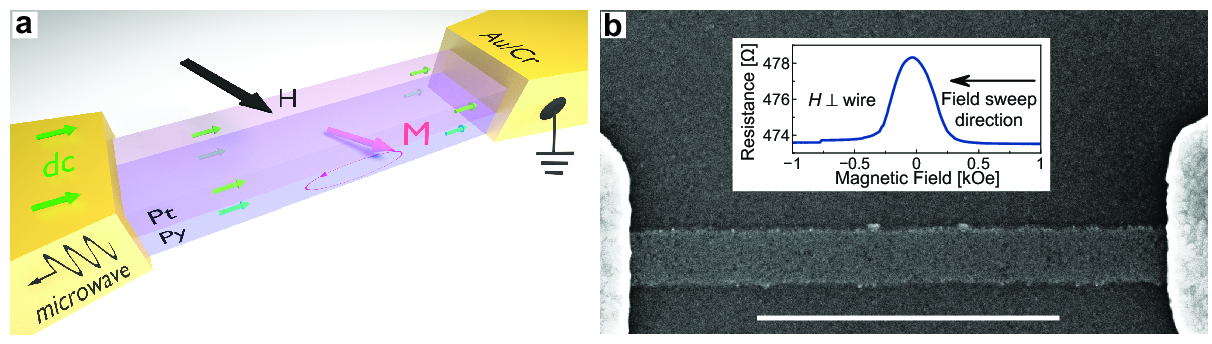}
\caption{\label{fig:device}\textbf{$|$ Sample structure.} (\textbf{a}) Schematic of a Pt/Py nanowire STO device: external magnetic field is shown as a black arrow, precessing Py magnetization is shown as a red arrow,  green arrows indicate the flow of direct electric current applied to the nanowire, microwave voltage generated by the sample is depicted as a wave with an arrow. (\textbf{b}) Scanning electron micrograph of the Pt/Py nanowire STO. The Pt/Py nanowire is the gray strip in the lower part of the image. The bright areas on both sides of the image are Au/Cr leads. Scale bar: 1 $\mu$m. The inset shows resistance versus in-plane magnetic field applied perpendicular to the nanowire measured at the bath temperature $T_\text{b}$ = 4.2 K and a bias current of 0.5 mA. The black arrow shows the magnetic field sweep direction.}
\end{figure}

\section*{Results} 

\subsection*{Sample description} The nanowire STO devices studied in this work are patterned from Pt(5 nm)/Py$\equiv$Ni$_{80}$Fe$_{20}$ (5 nm)/AlO$_x$(4 nm)/(GaAs substrate) multilayers deposited by magnetron sputtering. Multilayer nanowires that are 6 $\mu$m long and 190 nm wide are defined via e-beam lithography and Ar plasma etching as described in the Methods section. Two Au(35 nm)/Cr(7 nm) leads are attached to each nanowire with a 1.8 $\mu$m gap between the leads, which defines the active region of the device as shown in Fig. \ref{fig:device}.  The resistance of the device measured at bath temperature $T_\text{b}$ = 4.2 K versus magnetic field applied in the sample plane perpendicular to the nanowire is shown in the inset of Fig. \ref{fig:device}. This plot reveals that the anisotropic magnetoresistance (AMR) of the Pt/Py bilayer is 1 $\%$. 

\begin{figure}
\includegraphics[width=\linewidth]{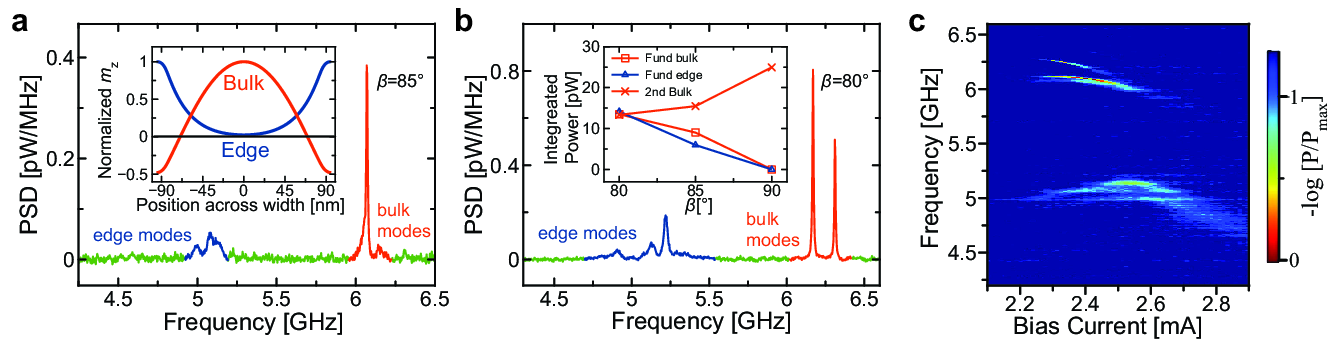}
\caption{\label{fig:STO}\textbf{$|$ Microwave emission spectra from the Pt/Py nanowire spin torque oscillator.} Power spectral density (PSD) of the microwave signal emitted by the Pt/Py nanowire at direct current bias $I_\text{dc}=$ 2.45 mA, bath temperature $T_\text{b}=$ 4.2 K and magnetic field $H=$ 890 Oe applied in the plane of the sample at an angle (\textbf{a}) $\beta=$ 85$^\circ$ and (\textbf{b}) $\beta=$ 80$^\circ$ with respect to the nanowire axis. The red, blue and green sections of the PSD curves correspond to the bulk spin wave modes, edge spin wave modes and the baseline respectively. (\textbf{c}) Dependence of the emission spectrum on $I_\text{dc}$ for $H=$ 890 Oe and $\beta=$ 85$^\circ$. The inset in (a) shows the spatial profiles of the edge (blue) and bulk (red) spin wave modes across the nanowire width given by micromagnetic simulations. The inset in (b) shows angular ($\beta$) dependence of the integrated power in the fundamental harmonic of the bulk group of spectral peaks (red squares), the second harmonic of the bulk group of peaks (red crosses), and the fundamental harmonic of the edge group of peaks (blue triangles) measured at $I_\text{dc}=$ 2.4125 mA and $H=$ 890 Oe. Lines are guides to the eye.}
\end{figure}

\subsection*{Electrical  measurements} To study self-oscillatory magnetic dynamics excited by SO torques, we apply a saturating magnetic field ($H >$ 0.5 kOe) in the plane of the sample in a direction nearly perpendicular to the nanowire axis. In this configuration, SO torques applied to the Py magnetization act as effective magnetic damping \cite{Ando2008}. We apply a direct current bias $I_\text{dc}$ to the nanowire and measure the microwave signal emitted by the device using a spectrum analyzer \cite{Kiselev2003}. The microwave signal $V_\text{ac}$ $\sim I_\text{dc} \delta R_\text{ac}$ is generated by the AMR resistance oscillations $\delta R_\text{ac}$ arising from the magnetization self-oscillations \cite{Liu2013}.  Microwave signal emission, shown in Fig. \ref{fig:STO}, is only observed above a critical current $I_\text{c}$ with the current polarity corresponding to SO torques acting as negative damping \cite{Ando2008}. We measured the microwave emission for five nominally identical devices and found similar results for all these samples. 

For all samples, the microwave emission spectra for $I_\text{dc}>I_\text{c}$ exhibit two groups of closely spaced peaks with a frequency gap between the groups of $\sim$1 GHz. Each group consists of 1 to 4 distinct emission peaks separated from each other by tens to hundreds of MHz. The high and low frequency groups of peaks appear at similar critical currents. Both groups of peaks are observed in the entire range of magnetic fields ($H=$ 0.5 -- 1.5 kOe) employed in this study. For the high frequency group of peaks, microwave emission is observed not only at the fundamental frequency shown in Fig. \ref{fig:STO}, but also at the second harmonic. As illustrated in the inset of Fig. \ref{fig:STO}b, the emitted power at the fundamental frequency is zero for magnetic field applied at an angle $\beta=$ 90$^\circ$ with respect to the nanowire axis and increases with decreasing $\beta$. In contrast, the emitted power in the second harmonic has a maximum at $\beta=$ 90$^\circ$ and decreases with decreasing $\beta$. Such angular dependence of the emitted power in the fundamental and second harmonic is expected for a microwave signal arising from AMR. For the low frequency group of peaks, no emission is seen at the second harmonic, which can be explained by the smaller amplitude of magnetization precession reached by these modes and an equilibrium magnetization direction within the mode excitation area being closer to the nanowire axis. Fig. \ref{fig:STO}c illustrates the dependence of the emission spectra on $I_\text{dc}$ for $H = 890$ Oe and $\beta = 85^\circ$. 

\begin{figure}
\includegraphics[width=\linewidth]{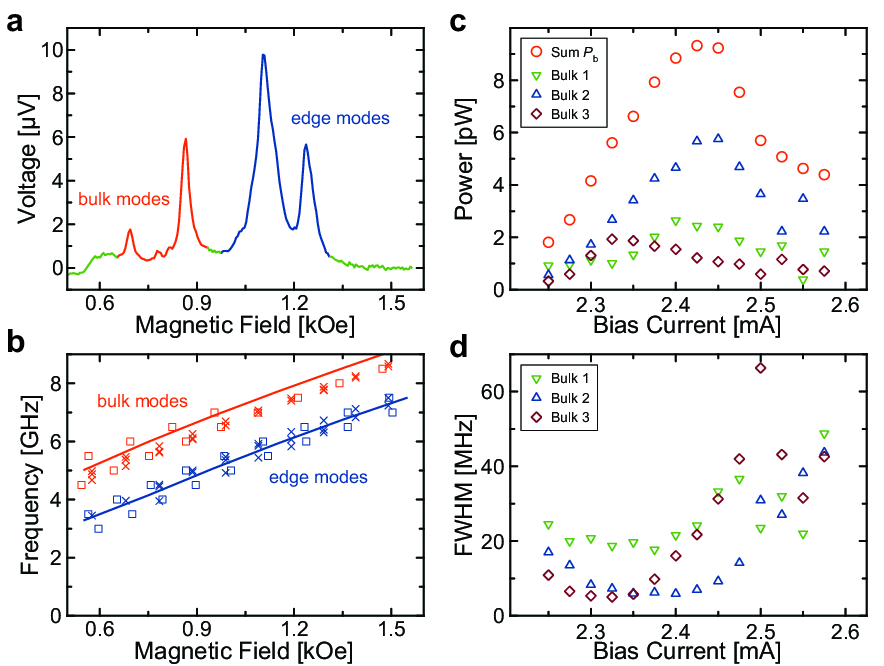}
\caption{\label{fig:FMR}\textbf{$|$ Spin wave modes of the STO.} (\textbf{a}) ST-FMR spectrum of the nanowire device measured at the microwave drive frequency of 6 GHz, $\beta=$ 85$^\circ$ and $I_\text{dc}=$ 2.0 mA $<I_\text{c}$. The red, blue and green sections of the curve represent the bulk modes, edge modes and the baseline respectively. (\textbf{b}) Frequency versus magnetic field applied at $\beta=$ 85$^\circ$: (squares) spin wave eigenmodes measured by ST-FMR, (crosses) self-oscillatory modes at $I_\text{c}$, and (lines) bulk and edge spin wave eigenmodes given by micromagnetic simulations for an ideal nanowire. Color scheme: red and blue represent bulk and edge spin wave modes respectively. (\textbf{c}) Bias current dependence of the integrated emitted power in three individual peaks of the bulk group of self-oscillatory modes as well as the sum of integrated powers of all bulk modes $P_\text{b}$. (\textbf{d}) Bias current dependence of the spectral linewidths (FWHM) of the individual peaks in the bulk group. Symbols in (c) and (d): (green triangles) bulk mode 1, (blue triangles) bulk mode 2, (purple diamonds) bulk mode 3, and (red circles) the total power of three bulk modes.}
\end{figure}

In order to determine the origin of the microwave emission signals, we make measurements of the spectrum of spin wave eigenmodes of the nanowire using spin torque ferromagnetic resonance (ST-FMR) \cite{Tulapurkar2005,Sankey2006,Liu2011}. In this technique, a microwave current $I_\text{ac}$ applied to the nanowire excites magnetization dynamics in Py by the combined action of current-induced SO torques and the Oersted field from the current in Pt, and thereby generates AMR resistance oscillations at the frequency of the microwave drive \cite{Liu2011}. Mixing of the current and resistance oscillations as well as variation of the time-averaged sample resistance in response to the microwave drive \cite{Wang2009, Mecking2007} give rise to a direct voltage $V_\text{dc}$ that is measured as a function of magnetic field $H$ applied to the sample. Peaks in $V_\text{dc}(H)$ arise from resonant excitation of spin wave eigenmodes of the nanowire. An ST-FMR spectrum of spin wave eigenmodes measured at $\beta = 85^\circ$, drive frequency of 6 GHz and $I_\text{dc}=$ 2.0 mA $< I_\text{c}$ is shown in Fig. \ref{fig:FMR}a. 

Similar to the microwave emission spectra, we observe two groups of modes in the ST-FMR spectra. In Fig. \ref{fig:FMR}b, we compare the field dependence of the eigenmode frequencies measured by ST-FMR at $I_\text{dc}<I_\text{c}$ to the frequencies of self-oscillatory modes measured at $I_\text{c}$. This figure demonstrates that the frequencies of all self-oscillatory modes at $I_\text{c}$ coincide with the frequencies of spin wave eigenmodes of the system measured by ST-FMR. Therefore, all self-oscillatory modes of the system arise directly from spin wave eigenmodes of the nanowire. This type of eigenmode self-oscillation is qualitatively different from the spin wave bullet mode excited by SO torques in a planar point contact to an extended ferromagnetic film \cite{Demidov2012}. The bullet mode is a nonlinear type of oscillation, self-localized to a region with dimensions below 100 nm and with frequency below the spectrum of spin wave eigenmodes of the film \cite{Slavin2005}. 

It is well known that spin wave eigenmodes of a transversely magnetized thin-film ferromagnetic nanowire can be classified as bulk and edge eigenmodes \cite{Zheng2014, Bayer2006, Park2002, McMichael2006}. These eigenmodes have spatially inhomogeneous profiles along the wire width with reduced (enhanced) amplitude near the wire edges for the bulk (edge) eigenmodes. The frequencies of the edge spin wave modes lie below those of the bulk modes due to reduced internal magnetic field near the wire edges \cite{Bayer2006}. The frequencies of all eigenmodes are sensitive to the values of magnetic parameters of the Py film, which are different from their bulk values due to the influence of  proximate nonmagnetic layers. In addition, the frequencies of the edge modes depend on the edge roughness and spatial variation of the film magnetic properties (magnetic dilution) at the nanowire edges induced by etching \cite{McMichael2006}. 

In the Supplementary Note, we describe the micromagnetic \cite{Donahue1999} fitting of three sets of experimental data: wire resistance versus in-plane magnetic field, wire resistance versus out-of-plane magnetic field and quasi-uniform spin wave mode frequency versus magnetic field applied parallel to the wire axis. These results can be used to determine the values of saturation magnetization $M_\text{s}$= 608 emu cm$^{-3}$, surface magnetic anisotropy $K_\text{s}$ = 0.237 erg cm$^{-2}$ and edge dilution depth $D$ = 10 nm, where $D$ is defined as the distance from the wire edge over which magnetization linearly increases from zero to the full film value \cite{McMichael2006}. These values of $M_\text{s}$, $K_\text{s}$ and $D$ are consistent with previous studies of thin Py nanomagnets \cite{Krivorotov2004,Rantschler2005,Nembach2013}. Using these values of the nanowire magnetic parameters and assuming translational invariance along the nanowire axis, we perform micromagnetic simulations  to find the spectrum of spin wave eigenmodes for magnetic field applied in the sample plane at $\beta=$ 85$^\circ$. We find that the two lowest frequency spin wave modes are the edge and bulk modes, whose spatial profiles (defined as normalized out-of-plane amplitude of dynamic magnetization $m_\text{z}$) across the wire width are shown in the inset of Fig. \ref{fig:STO}a. The frequencies of these two modes versus in-plane magnetic field are shown in Fig. \ref{fig:FMR}b. 

It is clear from Fig. \ref{fig:FMR}b that the frequencies of the calculated edge (bulk) modes are similar to the low (high) frequency groups of eigenmodes measured by ST-FMR and observed in STO measurements. We thus conclude that the high (low) frequency groups of experimentally observed eigenmodes are closely related to the bulk (edge) eigenmodes of an ideal nanowire with translational invariance along the nanowire axis. As demonstrated by our micro-focus Brillouin light scattering (BLS) measurements described below, the fine splitting within a group originates from the spin wave spectrum quantization along the nanowire length. Such quantization can be  imposed by reflection of spin waves at the edge of the Au/Cr leads, where the effective magnetic damping rapidly varies as a function of position along the wire. Fig. \ref{fig:FMR}c shows the dependence of the integrated microwave power emitted by the STO device on direct bias current $I_\text{dc}$ for the three spectral peaks comprising the bulk group. The sum of the integrated powers in the three peaks versus $I_\text{dc}$ is shown as well. Although the power in each individual peak exhibits irregularities as a function of $I_\text{dc}$, the sum of integrated powers of the entire group of bulk peaks is a smooth function of $I_\text{dc}$. The same trend is found for the edge group of peaks. As discussed in the Methods section, the maximum integrated power emitted by the bulk mode corresponds to the sample resistance oscillations that are 15$\%$ of the full AMR amplitude. Similar maximum amplitude of resistance oscillations is found for the edge mode. This demonstrates that large-amplitude magnetization oscillations can be excited by SO torques in both the bulk and the edge modes of the nanowire. Fig. \ref{fig:FMR}d shows the full width at half maximum (FWHM) of three spectral peaks of the bulk group versus the bias current. The minimum FWHM for each of the peaks is observed near the current value corresponding to the maximum integrated power of the peak, which is typical for STO dynamics \cite{Slavin2009}.

\begin{figure}
\includegraphics[width=\linewidth]{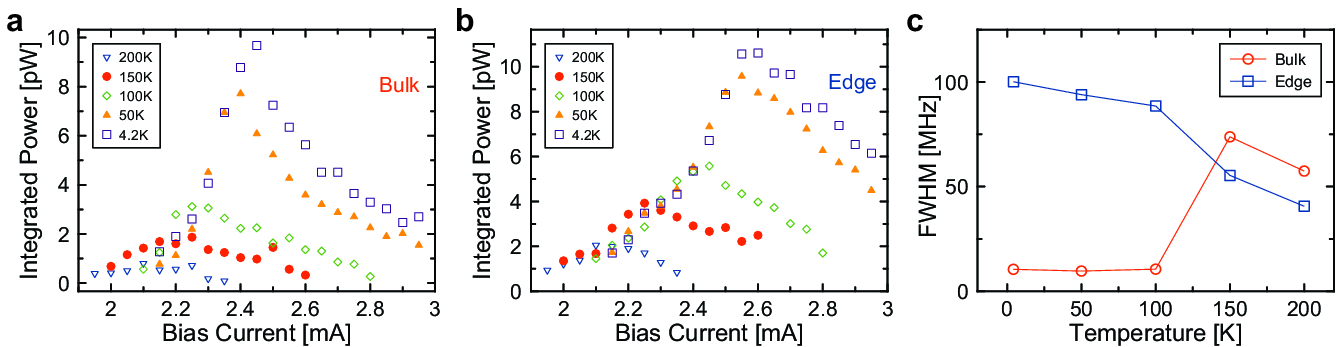}
\caption{\label{fig:Tdep}\textbf{$|$ Temperature dependence of the emission spectra.} Bias current dependence of the integrated microwave power emitted by the bulk (\textbf{a}) and edge (\textbf{b}) spin wave modes measured at $\beta=$ 85$^\circ$, $H=$ 890 Oe and several values of the bath temperature $T_\text{b}=$ 200 K (blue triangles), 150 K (red dots), 100 K (green diamonds), 50 K (yellow triangles) and 4.2 K (purple squares). (\textbf{c}) Temperature dependence of the spectral linewidth (FWHM) of the highest-power bulk (red circles) and edge (blue squares) self-oscillatory modes measured at the bias current of the maximum integrated power of the mode.}
\end{figure}

We also make measurements of the microwave signal emission as a function of temperature. Figs. \ref{fig:Tdep}a and \ref{fig:Tdep}b show the dependence of the total integrated power in the bulk and edge groups of peaks measured at several values of the bath temperature $T_\text{b}$. The integrated power decreases with increasing $T_\text{b}$ and vanishes at $T_\text{b} \approx$ 250 K. The actual temperature of the nanowire is significantly higher than $T_\text{b}$ due to Ohmic heating. The nanowire temperature can be estimated from measurements of the nanowire resistance as functions of $T_\text{b}$ and $I_\text{dc}$ \cite{Liu2013}. As described in the Methods section, the actual nanowire temperature at $I_\text{c}$ and $T_\text{b} =$ 4.2 K is approximately 150 K. Fig. \ref{fig:Tdep}c illustrates the temperature dependence of the nanowire STO spectral linewidth. This figure shows FWHM of the dominant (highest integrated power) bulk and edge peaks versus temperature. For each peak, the FWHM is measured at the current bias that maximizes the integrated power of the peak. The spectral linewidth of the edge peak decreases with increasing temperature, which cannot be explained by a single-mode STO theory \cite{Slavin2009}. However, such temperature dependence has been previously observed for multi-mode STOs and was successfully explained by an STO theory taking into account coupling between multiple self-oscillatory modes \cite{Muduli2012}. The linewidth of the dominant bulk peak is an order of magnitude smaller than that of the dominant edge peak for $T_\text{b}\leq$100 K but it rises precipitously to a value similar to that of the edge peak for $T_\text{b}>$100 K, which might result from enhanced coupling between the bulk and the edge modes for $T_\text{b}>$100 K. 

\subsection*{Brillouin Light Scattering measurements} In order to better understand the nature of the self-oscillatory dynamics induced by SO torques in the Pt/Py nanowire system and to directly confirm that the self-oscillatory modes occupy the entire active region of the nanowire, we make BLS measurements \cite{Demokritov2008} of the current-driven magnetization dynamics in this system. Since a 5 nm thick layer of Pt is not sufficiently transparent for BLS studies, we make a separate batch of samples for BLS measurements with the reverse order of deposition of the AlO$_x$, Py and Pt layers. These AlO$_x$(2 nm)/Py(5 nm)/Pt(7 nm)/(sapphire substrate) nanowire samples are prepared by e-beam lithography and liftoff technique as described in the Methods section. The microwave signal emission from these samples is similar to that of the samples in Figs. 1--4, with two notable differences: the amplitude of self-oscillations of the edge mode is significantly higher than that of the bulk mode and the microwave emission from the edge mode persist up to $T_\text{b}=$ 300 K.  These differences are likely to originate from the different sample fabrication procedure: the lift-off fabrication process creates less nanowire edge damage than the Ar plasma etching process. 

\begin{figure}
\includegraphics[width=0.85\linewidth]{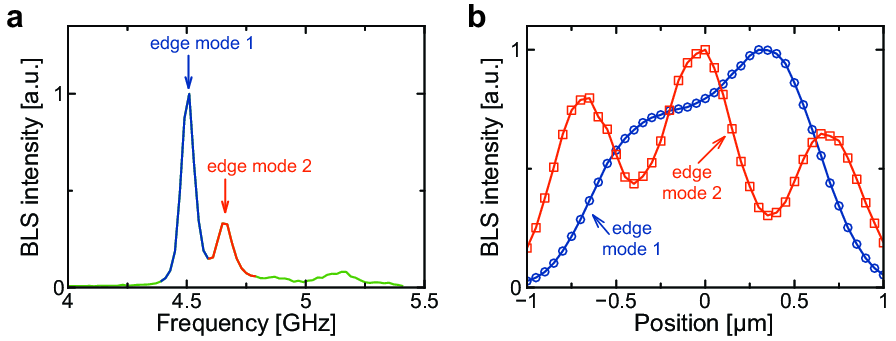}
\caption{\label{fig:BLS}\textbf{$|$ BLS characterization of the auto-oscillating modes.} (\textbf{a}) BLS spectrum acquired by placing the probing laser spot at the center of the nanowire. BLS intensity is proportional to the intensity of the dynamic magnetization. The blue, red and green sections of the curve represent the edge mode 1, edge mode 2 and the baseline respectively. (\textbf{b}) Spatial profiles of the intensity of the dynamic magnetization in the section parallel to the nanowire axis: (blue circles) edge mode 1 and (red squares) edge mode 2. The data were obtained at $H =$ 550 Oe and the bias current $I_\text{dc}=$ 2.4 mA.}
\end{figure}

The main results of the BLS measurements are presented in Fig. \ref{fig:BLS}. Figure \ref{fig:BLS}a shows a representative BLS spectrum measured by placing the probing laser spot at the center of the nanowire. In agreement with the results of electronic measurements of samples with Pt on top, the BLS spectrum exhibits a series of auto-oscillation peaks with the typical frequency separation of hundreds of MHz, which belong to the group of edge modes. By using the high spatial resolution of the BLS measurements, we identify the individual auto-oscillation peaks within the group. For this we fix the BLS detection frequency at the frequency of one of the peaks and record spatial profiles of the dynamic magnetization by moving the probing laser spot along the nanowire axis with the spatial step size of 50 nm. As seen in Fig. \ref{fig:BLS}b, the spatial profiles corresponding to the two peaks are fundamentally different. While the profile for the first edge mode exhibits a slightly distorted bell-like shape, the profile for the second peak possesses three maxima. Based on the obtained data we conclude that the individual auto-oscillation peaks within the groups correspond to the standing-wave modes quantized in the direction parallel to the nanowire axis \cite{Ulrichs2011}. The first peak corresponds to the combination of the fundamental mode having no nodes and the antisymmetric mode possessing one nodal line at the center. These two modes are indistinguishable in the BLS spectrum, likely due to their small frequency separation. The second peak apparently corresponds to the mode possessing two nodal lines at the positions of the minima of the measured profile. We note that, since the BLS technique is sensitive to the intensity of the dynamic magnetization, the measured profile does not show a change of the sign across the positions of the nodal lines.

\section*{Discussion} Recent experiments \cite{Demidov2011} demonstrated that application of spatially uniform SO torques to an extended ferromagnetic film does not result in excitation of magnetic self-oscillations because the amplitudes of all spin wave modes of the film are limited by nonlinear magnon scattering processes. Therefore, our observation of self-oscillatory dynamics excited by SO torques in the entire 1.8 $\mu$m long active region of a ferromagnetic nanowire is surprising. We argue that quantization of the spin wave spectrum in the nanowire geometry reduces the number of available nonlinear magnon scattering channels and thereby enables excitation of self-oscillatory dynamics of the low-energy spin wave eigenmodes of the nanowire. 

\begin{figure}
\includegraphics[width=\linewidth]{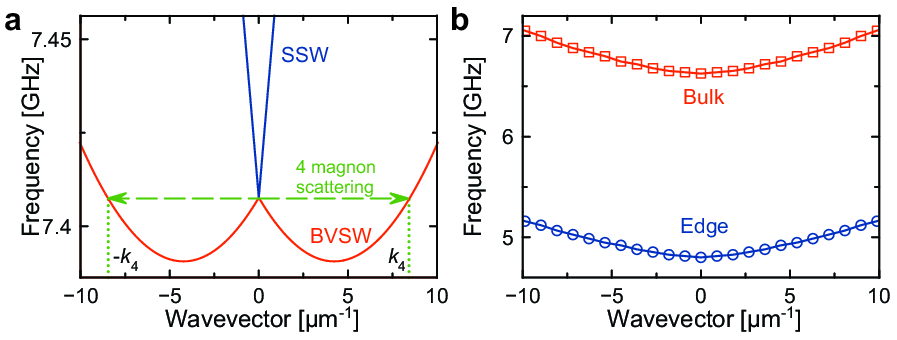}
\caption{\label{fig:Disp}\textbf{$|$ Py thin film and nanowire spin wave dispersion.} (\textbf{a}) Spin wave dispersion relation for BVSW (red) and SSW (blue) modes of a 5 nm thick Py film at $H=$ 890 Oe \cite{DeWames1970}. Arrows indicate energy- and momentum-conserving four-magnon scattering of two uniform mode magnons into two BVSW magnons with wave vectors $k_4$ and $-k_4$. (\textbf{b}) Spin wave dispersion relation of the 190 nm wide nanowire numerically calculated for $\beta=$ 85$^\circ$ and $H=$ 890 Oe. The red squares and blue circles indicate the bulk and edge modes respectively. Lines are guides to the eye. Energy- and momentum-conserving four-magnon scattering is not allowed within both the bulk and the edge mode branches of the nanowire spin wave dispersion.}
\end{figure}

Fig. \ref{fig:Disp}a shows the spin wave dispersion relation of a 5 nm thick extended Py film calculated for spin wave propagating parallel (backward volume spin waves, BVSW) and perpendicular (surface spin waves, SSW) to the in-plane magnetization vector \cite{DeWames1970}. Energy- and momentum-conserving four-magnon scattering processes from the uniform ($\textbf{k}=0$) mode into spin wave modes with a finite wave vector $\textbf{k}_4$ are allowed for the BVSW but not for the SSW modes as schematically illustrated in Fig. \ref{fig:Disp}a. This nonlinear scattering channel is always present in the 2D film geometry and it contributes to the nonlinear damping of the uniform mode. When such a film is patterned into a nanowire aligned perpendicular to the magnetization direction, the BVSW mode spectrum becomes quantized and the energy-conserving four-magnon scattering channel becomes suppressed for wires narrower than approximately $\pi k_4^{-1}$, which is $\approx 0.5$ $\mu$m for our 5 nm thick Py nanowires. Fig. \ref{fig:Disp}b shows the spin wave dispersion relation numerically calculated for our 190 nm wide nanowire sample as described in the Supplementary Note. The frequencies of the bulk and edge modes are shifted below the frequencies of spin waves in the extended film shown in Fig. \ref{fig:Disp}a by the demagnetizing field of the nanowire. The dispersion branches of both the edge and the bulk spin wave modes do not have minima at $k\neq$0 and therefore do not support the energy- and momentum-conserving four-magnon scattering within each of the branches. 

Since some nonlinear magnon scattering processes (such as the four-magnon process) responsible for limiting the amplitude of low-frequency spin wave modes in the 2D film are eliminated in the nanowire, ST excitation of large-amplitude self-oscillatory spin wave modes may become possible above the critical current. At higher currents, SO torques increase the occupation numbers for all magnon modes, which results in enhanced scattering in the remaining nonlinear channels and the associated decrease of the amplitude of self-oscillations seen in Fig. \ref{fig:FMR}c as well as in the reduction of the saturation magnetization and the associated decrease of the self-oscillatory mode frequency with current seen in Fig. \ref{fig:STO}c. Our data in Fig. \ref{fig:Tdep} showing decreasing self-oscillation amplitude with increasing temperature suggest that enhanced population of thermal magnons and magnon-phonon scattering might also play a role in suppression of the self-oscillatory dynamics. Development of a detailed theory accounting for all magnon scattering channels in the nanowire in the presence of SO torques is required for quantitative explanation of our experimental data, and we hope that our work will stimulate the development of such a theory. 

In conclusion, we demonstrated that spatially uniform spin torques can excite self-oscillations of magnetization in a 1D ferromagnetic system -- a Py nanowire. The self-oscillatory modes induced by spin torque in this system directly arise from the bulk and edge spin wave eigenmodes of the nanowire. The 1D nanowire geometry offers unique advantages for studies of magnetization dynamics driven by spin torques over the 2D extended thin film system, in which spatially uniform spin torque does not excite magnetization self-oscillations. Our results suggest that the self-oscillatory dynamics in the nanowire geometry is enabled by geometric confinement of magnons that suppresses nonlinear magnon scattering. Our work demonstrates the feasibility of spin torque oscillators with the active region dimensions extended beyond the nanometer length scale.

\section*{Methods}

\subsection*{Sample fabrication} All layers of the Pt(5 nm)/Py(5 nm)/AlO$_x$(4 nm)/(GaAs substrate) samples are deposited by magnetron sputtering at room temperature. The 6 $\mu$m long nanowires are defined via e-beam lithography and Ar plasma etching using e-beam evaporated Cr mask. The Au(35 nm)/Cr(7 nm) leads with a 1.8 $\mu$m gap between them defining the active region of the nanowire STO are made via e-beam lithography and e-beam evaporation of the Au/Cr bilayer followed by lift-off. The fabrication process of the AlOx(2 nm)/Py(5 nm)/Pt(7 nm)/(c-plane sapphire substrate) samples for the micro-BLS measurements starts with the deposition of a 5 nm thick Pt layer onto a sapphire substrate at 585 $^\circ$C and annealing for 1 hour at the same temperature, which results in growth of a continuous Pt film as verified by high resolution SEM and atomic force microscopy imaging. Then the nanowire is defined on top of the Pt film via e-beam lithography, brief Ar plasma cleaning immediately followed by \textit{in situ} room temperature sputter deposition of AlOx(2 nm)/Py(5 nm)/Pt(2 nm) trilayer, and lift-off. The Au(35 nm)/Cr(7 nm) leads are defined via e-beam lithography and e-beam evaporation of the Au/Cr bilayer followed by lift-off. At the final fabrication step, Ar plasma etching is used to remove the 5 nm thick bottom Pt layer everywhere but under the Py nanowire and the Au/Cr leads.

\subsection*{Brillouin Light Scattering measurements} Micro-focus BLS measurements were performed at room temperature by focusing light produced by a continuous-wave single-frequency laser operating at a wavelength of 532 nm into a diffraction-limited spot. The light scattered from magnetic oscillations was analyzed by a six-pass Fabry-Perot interferometer TFP-1 (JRS Scientific Instruments, Switzerland) to obtain information about the BLS intensity proportional to the square of the amplitude of the dynamic magnetization at the location of the probing spot.

\subsection*{Estimate of the amplitude of resistance self-oscillations} In this section, we estimate the maximum amplitude of the resistance oscillations achieved by the bulk and edge modes in the self-oscillatory regime. The measured integrated power at the fundamental frequency emitted by the bulk group of modes $P_\text{b}$ (corrected for frequency-dependent attenuation and amplification in the measurement circuit) is plotted as a function of direct bias current $I_\text{dc}$ in Fig. 3c. This power is directly related to the amplitude of the nanowire resistance oscillations $\delta R_\text{ac}$ at the fundamental frequency of magnetization self-oscillations. These resistance oscillations arise from anisotropic magneto-resistance (AMR) of the nanowire. The microwave voltage generated by the nanowire STO device at the fundamental frequency of the bulk mode $I_\text{dc} \delta R_\text{ac}$ is detected by a 50 Ohm microwave spectrum analyzer as microwave power $P_\text{b}$ \cite{Kiselev2003}:
\begin{equation} \label{eqn:Pb}
P_\text{b}=\frac{1}{2 R_{50}}\left( I_\text{dc} \delta R_\text{ac} \frac{R_{50}}{R+R_{50}} \right)^2
\end{equation}

where $R_{50}\equiv50$ Ohm is the spectrum analyzer impedance and $R$ is the nanowire resistance. From this equation, we can express the amplitude of resistance oscillations at the fundamental frequency as a function of the emitted power:
\begin{equation} \label{eqn:Rac}
\delta R_\text{ac}=\frac{R+R_{50}}{I_\text{dc} \sqrt{R_{50}}} \sqrt{2 P_\text{b}}
\end{equation}

Using Eq. (\ref{eqn:Rac}) and the maximum experimentally measured integrated power in the bulk group of modes $P_\text{b}=$ 10 pW at $I_\text{dc}=$ 2.45 mA as shown in Fig. 4a, we evaluate $\delta R_\text{ac}\approx$ 0.15 Ohm. A similar maximum value of $\delta R_\text{ac}$ arising from the edge mode self-oscillations is calculated from the data in Fig. 4b. These values of $\delta R_\text{ac}$ are substantial fractions of the full AMR amplitude of 1 Ohm shown in the inset in Fig. 1b. This demonstrates that large amplitude of magnetization self-oscillations can be achieved by both the bulk and the edge modes of the nanowire STO.

\subsection*{Sample temperature} Direct current bias $I_\text{dc}$ can significantly increase the temperature of the nanowire due to Ohmic heating. In order to estimate the actual temperature of the nanowire at large $I_\text{dc}$, we compare measurements of the wire resistance as a function of temperature at small $I_\text{dc}=$ 0.1 mA to measurements of the wire resistance versus $I_\text{dc}$ taken at the bath temperature $T_\text{b}=$ 4.2 K \cite{Liu2013}. These data are plotted in Supplementary Fig. 4, which shows that resistance of the nanowire is approximately quadratic in both $T_\text{b}$ and $I_\text{dc}$. This allows us to estimate the actual temperature of the nanowire directly. For example, at $I_\text{dc}=$ 2.0 mA, the actual temperature of the nanowire $T \approx$ 150 K.


\section*{Acknowledgments} 
This work was supported by NSF Grants DMR-1210850, ECCS-1309416, and by the FAME Center, one of six centers of STARnet, an SRC program sponsored by MARCO and DARPA. Funding by the DFG/NSF in the framework of the Materials World Network program is acknowledged. Support of the program Megagrant № 2013-220-04-329 of the Russian Ministry of Education and Science is also acknowledged.

\section*{Author contributions}
Z.D., A.S., L.Y. and B.Y. designed and made the samples. Z.D., A.S. and L.Y. performed electrical characterization of the samples. A.S. performed micromagnetic simulations. V.E.D. performed micro-BLS measurements. J.L., S.O.D. and I.N.K. formulated the experimental approach and managed the project. All authors analyzed the data and co-wrote the paper. Z.D., A.S., L.Y. and B.Y. contributed equally to this work.

\section*{Additional information}
 Supplementary Information accompanies this paper at ... .
 
\section*{Competing financial interests} 
The authors declare no competing financial interests.

\section*{Correspondence}
 Correspondence and requests for materials should be addressed to Zheng Duan (email: Justin.Duan68@gmail.com).


\end{document}




\subsection*{Supplementary Note: Material parameters of the Permalloy nanowire from micromagnetic simulations}

Magnetic properties of the Py film in the Pt(5 nm)/Py(5 nm)/AlO$_x$(4 nm)/(GaAs substrate) nanowire samples are expected to differ from those of bulk Py. Previous studies of thin Py films interfaced with non-magnetic metallic layers and alumina have shown reduction of the saturation magnetization $M_\text{s}$ \cite{Krivorotov2004} and presence of perpendicular surface anisotropy $K_\text{s}$ \cite{Djemia1997,Rantschler2005}. In addition, patterning of the film into nanowires by Ar plasma etching creates material intermixing at the nanowire edges and results in magnetic edge dilution \cite{McMichael2006,Nembach2013}. Therefore, in order to be able to predict the spectrum of spin wave eigenmodes of the nanowire, we need to extract relevant magnetic parameters of the Py layer from experiment. In this section, we describe the micromagnetic \cite{Donahue1999} fitting of three sets of experimental data: wire resistance versus in-plane magnetic field, wire resistance versus out-of-plane magnetic field and quasi-uniform spin wave mode frequency versus magnetic field applied parallel to the wire axis. We explain how these fittings are used to determine approximate values of saturation magnetization $M_\text{s}$, surface magnetic anisotropy $K_\text{s}$ and the edge dilution depth $D$. 

The value of the exchange stiffness of bulk Py $A\approx 10^{-6}$ erg cm$^{-1}$ is known only approximately \cite{Maksymowicz1991, Smith1989} due to the lack of reliable direct methods for measuring this quantity. It is also known that the value of $A$ is usually reduced in thin films and nanomagnets compared to its bulk value \cite{Helmer2010}. In order to estimate the value of $A$ for the Py layer in our nanowire device, we employ the scaling relation $A \sim M_\text{s}^2$ and the value of saturation magnetization $M_\text{s}=$ 608 emu cm$^{-3}$ extracted from the micromagnetic fit described below. This scaling predicts the exchange stiffness to be approximately half of the bulk value, and thus we adopt $A=5\times10^{-7}$ erg cm$^{-1}$ throughout all micromagnetic simulations in this work. We also employ a thin-film value of the spectroscopic $g$-factor ($g$ = 2.03) measured for a 5 nm thick Py film \cite{Nibarger2003}. For quantification of the magnetic edge dilution, we adopt the model developed in Ref. \citenum{McMichael2006}, in which the saturation magnetization varies linearly from zero at the wire edge to the full film value $M_\text{s}$ over the edge dilution depth $D$. 

\begin{figure} [t!]
\includegraphics[width=\linewidth]{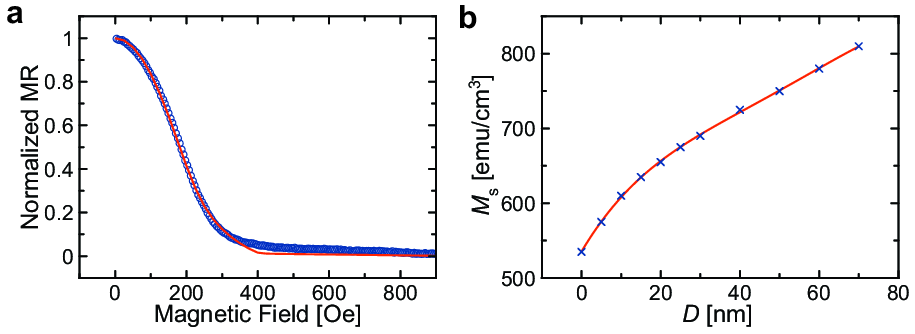}
\caption{\label{fig:MvsD}\textbf{Dilution depth dependence of $\mathbf{M_\text{s}}$.} (\textbf{a}) Normalized magnetoresistance of a 190-nm wide Pt(5 nm)/Py(5 nm) nanowire measured as a function of magnetic field applied in the plane of the sample perpendicular to the nanowire axis (blue circles). The solid red line shows a micromagnetic fit to the data for the edge dilution depth $D=$ 10 nm, which gives the value of $M_\text{s}=$ 608 emu cm$^{-3}$. (\textbf{b}) Blue crosses show the dependence of $M_\text{s}$ on $D$ given by the micromagnetic fit to the data in (a). The red line is a fourth order polynomial fit to the data.}
\end{figure}

We begin the micromagnetic fitting procedure by fitting the data of the normalized wire magnetoresistance $\hat{R}$ as a function of magnetic field $H_\text{ip}$ applied in the plane of the sample perpendicular to the wire axis (Supplementary Fig. \ref{fig:MvsD}a). This is a convenient starting point because $\hat{R}(H_\text{ip})$ is independent of $K_\text{s}$, which allows us to directly establish a relation between $M_\text{s}$ and $D$ through micromagnetic fitting to the $\hat{R}(H_\text{ip})$ data in Supplementary Fig. \ref{fig:MvsD}a. We divide the 190 nm wide, 5 nm thick Py nanowire into 2 nm $\times$ 1 nm $\times$ 5 nm micromagnetic cells and find the equilibrium configuration of magnetization for a given value of $H_\text{ip}$. We then calculate the wire resistance $R$ by dividing the wire into $N=190$  Py(5 nm)/Pt (5 nm) bilayer strips (each strip is 1 nm wide) along its length and use the equation for resistors connected in parallel to calculate the wire resistance $R=\left(\sum^{N}_{i=1} R_i^{-1}\right)^{-1}$. The resistance of each strip $R_i$ $(i=1..N)$ is calculated according to the AMR formula: $R_i=\tilde{R}_0+\Delta\tilde{R_i} \cos^2(\theta_i)$, where $\tilde{R}_0 = N R_0$ is the resistance of the Py(5 nm)/Pt (5 nm) bilayer strip magnetized perpendicular to the strip axis, $\Delta\tilde{R_i}$ is the full AMR of the $i$-th strip, $R_0$ is the resistance of the nanowire magnetized perpendicular to the nanowire axis $\Delta R=\left(\sum^{N}_{i=1}\left(N R_0+\Delta\tilde{R_i}\right)^{-1}\right)^{-1}-R_0$ is the full AMR of the nanowire and $\theta_i$ is the angle between magnetization in the $i$-th strip and the nanowire axis. This equation for calculating the wire resistance is a good approximation because $\Delta R \ll R_0$ and the resistivities of Py and Pt are similar to each other. We make the simplifying assumption that $\Delta\tilde{R_i}$ in the $i$-th strip is proportional to the value of $M_\text{s}$ in this strip \cite{Mcguire1975} so that AMR is reduced at the nanowire edges due to the magnetic dilution.

We choose several values of the edge dilution depth $D$ between 0 and 70 nm, and for each value of $D$ we perform a set of micromagnetic simulations to determine the value of $M_\text{s}$ that gives the best fit to the $\hat{R}(H_\text{ip})$ data. The resulting dependence of $M_\text{s}$ on $D$ is shown by circles in Supplementary Fig. \ref{fig:MvsD}b. We then perform a fourth order polynomial fit to the data in Supplementary Fig. \ref{fig:MvsD}b and thereby establish the functional dependence $M_\text{s}(D)$ that is used throughout the rest of our micromagnetic simulations.

\begin{figure}[t!]
\includegraphics[width=\linewidth]{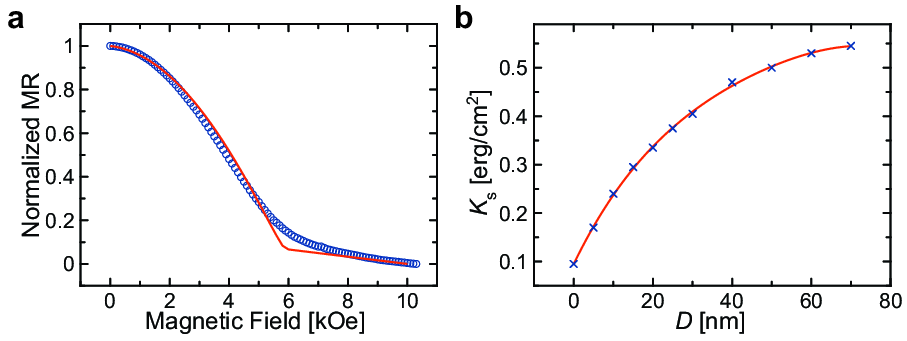}
\caption{\label{fig:KvsD}\textbf{Dilution depth dependence of anisotropy.} (\textbf{a}) Normalized magnetoresistance of a 190-nm wide Pt(5 nm)/Py(5 nm) nanowire measured as a function of the magnetic field applied normal to the sample plane (blue circles). The solid red line shows the micromagnetic fit to the data for the edge dilution depth $D=$ 10 nm, which gives the value of $K_\text{s}=$ 0.237 erg cm$^{-2}$. (\textbf{b}) Blue crosses show the dependence of $K_\text{s}$ on $D$ given by the micromagnetic fitting to the data in (a). The red line is a fourth order polynomial fit to the data.}
\end{figure}

We next fit the normalized nanowire magnetoresistance data measured as a function of magnetic field $H_\text{op}$ applied perpendicular to the sample plane (Supplementary Fig. \ref{fig:KvsD}a. We performed a set of micromagnetic simulations for several values of the edge dilution depth $D$ between 0 and 70 nm, and for each value of $D$, found the value of $K_\text{s}$ that gave the best micromagnetic fit to the $\hat{R}(H_\text{op})$ data in Supplementary Fig. \ref{fig:KvsD}a. In these simulations, we used the function $M_\text{s}(D)$ established in the $\hat{R}(H_\text{ip})$ fitting. The resulting dependence of $K_\text{s}$ on $D$ is shown by circles in Supplementary Fig. \ref{fig:KvsD}b. Fitting  a fourth order polynomial to the data in Supplementary Fig. \ref{fig:KvsD}b gives us the functional relation $K_\text{s}(D)$ that is used throughout the rest of our micromagnetic simulations. 

In order to determine the value of $D$ that best describes the properties of our nanowire device, we employ measurements of the quasi-uniform spin wave mode frequency $f_\text{q}$ as a function of magnetic field $H_{||}$ applied parallel to the nanowire axis. These measurements are made by the spin torque ferromagnetic resonance (ST-FMR) technique \cite{Tulapurkar2005,Sankey2006} described in the main text. The measured dependence of $f_\text{q}$ on $H_{||}$ is shown by crosses in Supplementary Fig. \ref{fig:fvsHeasyaxis}. To find the theoretical dependence $f_\text{q}(H_{||})$, we employ micromagnetic simulations to calculate the spectrum of spin wave eigenmodes of the nanowire. For these simulations, we simultaneously apply a pulse of SO torque and a pulse of Oersted field to the nanowire magnetization. Both pulses are generated by a spatially uniform sinc-shaped current pulse applied to the nanowire \cite{Venkat2013}:

\begin{figure}[t]
\includegraphics[width=0.6\linewidth]{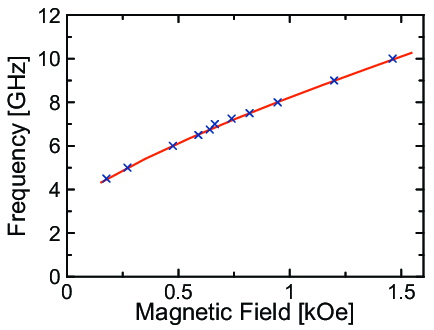}
\caption{\label{fig:fvsHeasyaxis} \textbf{Easy-axis field dependence of quasi-uniform mode resonance.} Frequency $f_\text{q}$ of the quasi-uniform spin wave mode of the nanowire as a function of magnetic field $H_{||}$ applied parallel to the nanowire axis. Blue crosses show the experimental data measured by ST-FMR, and the solid red line is the micromagnetic calculation result for the edge dilution depth $D=$ 10 nm.}
\end{figure}
\begin{equation} \label{eqn:sinc} 
I(t) = I_0\frac{\sin(2\pi f_\text{c}t)}{2 \pi f_\text{c} t}
\end{equation}

Such a current pulse excites all spin wave eigenmodes of the system that have amplitude profiles symmetric with respect to the nanowire center and have frequencies below the cutoff frequency $f_\text{c}=$ 20 GHz. The current pulse amplitude is chosen to be small enough that the magnetization dynamics remains in the linear regime. The magnetization of each micromagnetic cell is recorded as a function of time and then Fourier transformed. Peaks in the Fourier transform amplitude of the dynamic magnetization correspond to frequencies of spin wave eigenmodes of the nanowire. For a magnetic field $H_{||}$  applied along the length of the nanowire, our simulations show that the lowest frequency mode is the quasi-uniform spin wave mode without nodes along the wire width. Our simulations show that the frequency of this mode depends on the edge dilution depth $D$ and therefore we can use the experimental data in Supplementary Fig. \ref{fig:fvsHeasyaxis} to determine the value of $D$ appropriate for our nanowire device. Using the functions $M_\text{s}(D)$ and $K_\text{s}(D)$ established by our magnetoresistance data fitting, we perform a single parameter ($D$) fit of the data in Supplementary Fig. \ref{fig:fvsHeasyaxis} and find the best agreement with the data for $D=$ 10 nm. The solid line in Supplementary Fig. \ref{fig:fvsHeasyaxis} shows the best fit to the data for $D=$ 10 nm. 

The micromagnetic fitting described above gives us the set of magnetic parameters accurately describing the magnetic properties of the nanowire device: $D=$ 10 nm, $M_\text{s}$(10 nm) = 608 emu cm$^{-3}$ and $K_\text{s}$(10 nm) = 0.237 erg cm$^{-2}$. Using these parameters, we performed micromagnetic simulations to determine the spectrum of the nanowire spin wave eigenmodes for magnetic field applied in the plane of the sample at the angle $\beta=85^\circ$ with respect to the nanowire axis. As in the easy-axis field case described above, we apply a pulse of SO torque and Oersted field generated by a spatially uniform sinc-shaped current pulse and determine spin wave eigenmode frequencies from positions of spectral peaks in the Fourier transform of time-dependent dynamic magnetization. We find that the lowest frequency mode is the edge eigenmode, which amplitude is maximum at the wire edge. Its spatial profile across the wire width is shown in the inset of Fig. 2a of the main text. The next lowest frequency mode has bulk character and exhibits maximum amplitude in the center of the wire as shown in the inset of Fig. 2a of the main text. Solid lines in Fig. 3b of the main text show the calculated dependence of frequencies of these two modes on the applied magnetic field.

In order to calculate the spin wave dispersion relation of the bulk and edge eigenmodes, we apply a sinc-shaped magnetic field pulse to a 50$\times$190 nm$^2$ region in the center of the nanowire in order to excite spin waves with frequencies up to the cutoff frequency and wave vectors up to the inverse length of the excitation region. The spin wave dispersion relation is then obtained as a two dimensional Fourier transform of the dynamic magnetization: the spatial dimension along the wire length and the time dimension \cite{Venkat2013}.

\begin{figure} [h]
\includegraphics[width=0.5\linewidth]{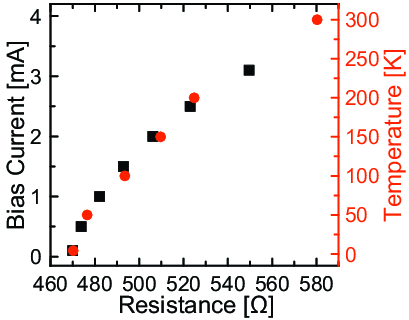}
\caption{\label{fig:RvsTandI} \textbf{Temperature and current dependence of resistance.} Resistance of the nanowire measured as a function of temperature at direct bias current $I_\text{dc}=$ 0.1 mA (red dots) and as a function of $I_\text{dc}$ at the bath temperature $T_\text{b}=$ 4.2 K (black squares).}
\end{figure}